# THERMIONIC SOURCES FOR ELECTRON COOLING AT IOTA

M. K. Bossard*, N. Banerjee, J. Brandt, Y.-K. Kim, University of Chicago, Chicago, IL, USA
M. Krieg, St. Olaf College, Northfield, MN, USA
B. Cathey, S. Nagaitsev, G. Stancari, Fermilab, Batavia, IL, USA

## Abstract

We are designing and fabricating two new thermionic sources of magnetized electrons for use in the electron lens project at the Integrable Optics Test Accelerator (IOTA) at Fermilab. These electron sources will be used for cooling of 2.5 MeV protons in the presence of intense space-charge. Furthermore, we are constructing an electron source test stand at the University of Chicago which will validate the electrical, thermal, and vacuum characteristics of thermionic sources. In this paper we present the progress made so far and the upcoming steps for the thermionic electron sources for electron cooling and the test stand.

## INTRODUCTION

Electron sources have played an important role in many areas throughout their rich history, such as the development of vacuum-tube technology, microscopy, modern electronic circuits, and particle accelerators [1]. Electron emission is the process by which electrons are emitted off of a material as their energies overcome the material's work-function [2]. There are many different physical stimuli which can supply the energy for electron emission, such as thermal-energy, photons, ion or electron bombardment, and high magnetic fields [3]. Thermionic electron sources are often used in electron lenses, which are flexible instruments for beam physics research and particle accelerator operations [4].

Electron lenses operate on the principle of a magnetically confined, low-energy electron beam overlapping with a re-circulating beam in a storage ring [4]. They are able to produce highly stable electron beams through the solenoidal magnetic field in the region of the beam. Electron lenses interact with the recirculating charged particles through electromagnetic fields allowing us to change the distribution of recirculating particles in phase-space. Example uses of electron lenses include space-charge compensation, beam-beam compensation [5, 6], halo reduction [7], and electron cooling.

The Integrable Optics Test Accelerator (IOTA) is a re-configurable 40 m storage ring at Fermilab [8, 9]. It acts as a test facility dedicated to research on intense beams, including the areas of Non-linear Integrable Optics (NIO) [10], space-charge [11], beam cooling [4], and single electron storage [12]. It can circulate both electrons and protons at kinetic energies up to 150 MeV and 2.5 MeV, respectively [11]. The layout of IOTA, along with the location of the electron cooler is shown in Fig. 1. In the ring, protons enter in section A. The beam circulates clockwise and co-propagates with the electrons in section DR.

___________________
* mbossard@uchicago.edu

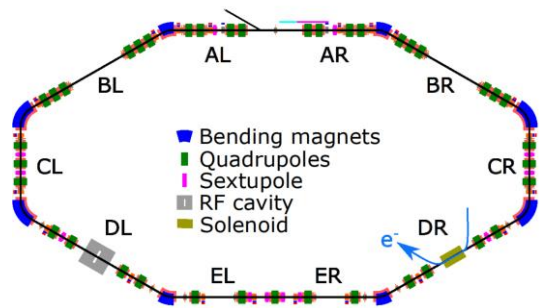

Figure 1: Schematic of the Integrable Optics Test Accelerator (IOTA). The ring is separated into sections, where the blue arrow in section DR represents the cool electron beam path.

Our goals in this project are to design, test, and commission two thermionic electron sources at IOTA. One source will be used as a tool for cooling proton beams with relatively small currents, to perform beam manipulations relevant for other experiments in IOTA. The other source will be used to investigate the dynamics and control of proton beams under intense space-charge [11].

In the upcoming sections, we will discuss electron cooling and its placement in IOTA, the cooler source designs, and the development of the test stand for the sources. We then will conclude and present the next steps.

## ELECTRON COOLING

In a recirculating ion beam's center of mass frame, the ions have random transverse velocities, corresponding to large emittance and large thermal energy. The thermal energy of a beam is increased by non-linear forces and interactions in the beam-line, such as space-charge effects, intra-beam scattering, and external non-linear magnetic fields throughout the accelerator path. To decrease the thermal energy of the ion beam, a cooling method must be employed.

In electron cooling, a beam of ions exchanges thermal energy with a co-propagating beam of electrons. As the beam of ions re-circulates through the accelerator, its thermal energy increases. To reduce this, it is periodically mixed with a bath of cold, low-energy electrons. The electrons move at the same average longitudinal velocity as the ions and the beams undergo Coulomb scattering, thus approaching thermal equilibrium and reducing the beam's thermal energy [13, 14].

Figure 2 shows the location of the electron cooler, zoomed into the DR section of the IOTA ring. It depicts the electron source, as well as the path that the electron beam takes. Solenoids are used throughout the electron beam propagation path to keep the beam magnetized. Figure 2 also shows the schematic of an electron source. The electrodes supply heat and potential differences needed for electron emission.

Table 1: Electron Cooler Parameters for IOTA

| Parameter | Values | | Unit |
|---|---|---|---|
| **Proton parameters** | | | |
| RMS Size ($\sigma_{b,x,y}$) | 3.22, 2.71 | | mm |
| **Main solenoid parameters** | | | |
| Magnetic field ($B_\parallel$) | 0.1 - 0.5 | | T |
| Length ($l_{cooler}$) | 0.7 | | m |
| Flatness ($B_\perp/B_\parallel$) | $2 \times 10^{-4}$ | | |
| **Electron parameters** | | | |
| Kinetic energy ($K_e$) | 1.36 | | keV |
| Temporal Profile | DC | | |
| Transverse Profile | Flat | | |
| Source temp. ($T_{cath}$) | 1400 | | K |
| Current ($I_e$) | 1 | 40 | mA |
| Radius ($a$) | 20 | 12 | mm |
| $\tau_{cool,x,y,s}$ | 37, 33, 18 | 3.4, 3.2, 1.7 | s |

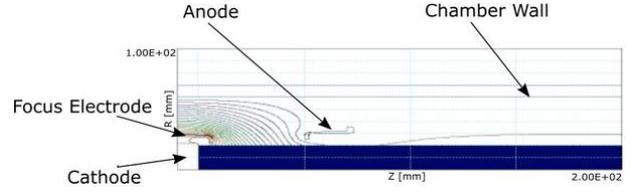

Figure 3: Present design of the basic thermionic source for electron cooling.

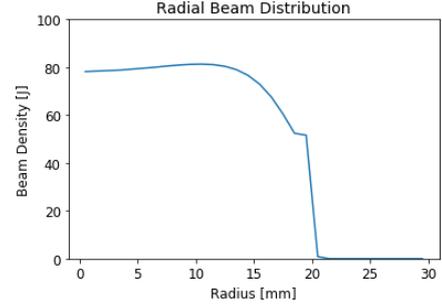

Figure 4: Beam distribution at 150 mm from the cathode for the basic thermionic source

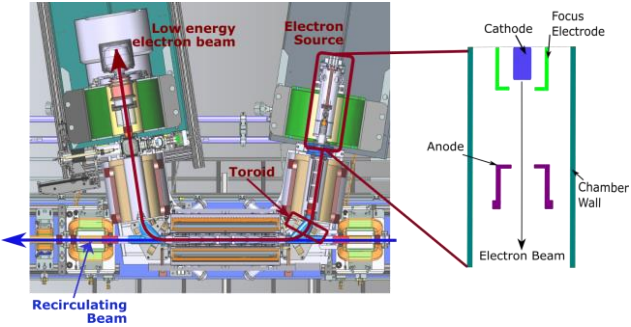

Figure 2: Electron lens setup at IOTA.

We will construct two electron coolers. Their parameters can be seen in Table 1. Again, one source will be used for basic cooling of proton beams with relatively small currents and perform beam manipulations relevant to other experiments in IOTA. The other source will be used to investigate the dynamics and control of proton beams under intense space-charge [11].

## ELECTRON SOURCE DESIGN

To create accurate designs for the thermionic sources, simulations are being developed. We are working with the simulation package TRAK, an advanced 2D code for charged-particle optics and particle source design [15]. The present design for the basic thermionic source in TRAK can be seen in Fig. 3. The main features of the electron source are depicted, where the entire system is cylindrically symmetric. The dark blue lines depict electron trajectories and the color gradient shows the electric field equipotentials with beam effects. The design of the cathode is a cylindrical metal with a flat emission surface. The focus electrode encircles the cathode and ensures that the emitting electrons do not diverge as they are emitted from the cathode. The anode is located further away and supplies an accelerating potential, forcing the electrons down the beam line. The entire system is simulated in a vacuum chamber with a magnetic field of 0.1 T, the same field as in the region of co-propagation, which constrains the motion of the electrons into helices, forcing the beam to remain collimated rather than diverging throughout its entire path. For the basic cooler design, as seen in Table 1, the potential difference between the cathode and the anode is 1.36 kV and the beam radius is 20 mm. The current emitted in this simulation is 90 mA. Therefore, the perveance of the electron source needs to be reduced to fulfill its function as a basic cooler.

The most relevant beam parameters for the source include the total current emitted, the radius of the beam, and the transverse beam current density distribution. An example of the current density distribution at 150 mm from the cathode can be seen in Fig. 4. As seen in this figure, the current density is approaching a flat distribution along the beam radius. In future simulations, we will explore optimizing the focus electrode and anode positions to obtain a better beam distribution and current.

Interestingly, the current density of the electron beam in our simulations is functional for a strong electron cooler. Therefore, not only do our electron source simulations approach a design for the basic cooler, they are currently producing a functional design for the strong cooler.

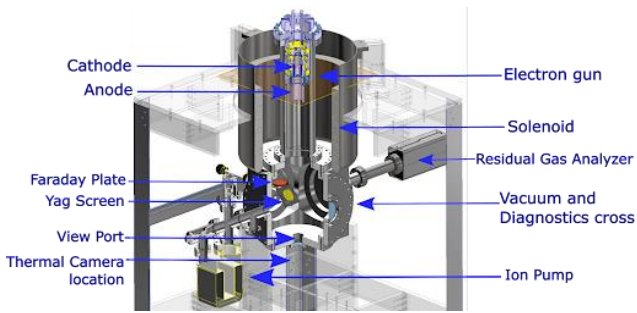

Figure 5: Test stand setup with an electron source in place.

## TEST STAND

Before commissioning the new electron sources at IOTA, we need to verify their operation at the University of Chicago. The design of a test stand for such verification is shown in Fig. 5. The electron beam flows downward, from the source into the diagnostics cube.

The cube consists of an active pumping system including turbomolecular and ion-getter pumps keeping the vacuum chamber below $10^{-7}$ mbar during electron source operation. Furthermore, to analyze the performance of the vacuum system there is a Residual Gas Analyzer (RGA). There also is a YAG:Ce screen, operating as a scintillator to analyze the beam distributions, as well as a Faraday plate to measure the current. The YAG:Ce screen and Faraday plate are attached to a movable arm, where they can individually lie under the beam at the correct extension. They can also move out of the beam's path so a viewport at the bottom of the cube can take images of the cathode. Below this viewport, a color CMOS camera will image the cathode and be used as a pyrometer. There will also be a camera outside the side viewport to image the YAG:Ce screen.

We recently have constructed the test stand. We also have designed and assembled the vacuum chamber as well as obtained an existing electron source from Fermilab for a thermal and vacuum test.

During the thermal and vacuum test, while the chamber remained at a pressure less than $10^{-7}$ mbar, the chamber pressure rose while turning on the current in the cathode filament, due to out-gassing of the filament and the rest of the chamber. We are currently working on baking the whole setup at 150°C to achieve a vacuum pressure less than $10^{-8}$ mbar when the cathode is cold.

A diagnostic we are developing is the measurement of the cathode temperature through the chamber's bottom viewport. We are developing a ratio pyrometer [16] to determine a body's temperature from an image taken by a color CMOS camera. In the future, we will improve this temperature algorithm method by further analyzing uncertainties and noise.

Other diagnostics we are developing include mounts for both of the cameras. We are also developing an interface to control and readback the camera, pytrometer, Faraday plate, magnet, and electron source filament and electrode parameters. The user interface we are using for this is based on Node-RED, a flow-based development tool.

## CONCLUSIONS AND NEXT STEPS

We are creating electron sources to cool proton beams and research electron cooling's influence on ion space-charge forces at high-intensity limits. To generate appropriate electron beams, we are designing and simulating two electron sources, where the designs of the sources will be altered to fit the needs at IOTA. We have changed electrode features in simulations, which approached the desired parameters for the basic electron source and produced the beam density for the strong electron source. Furthermore, we are working towards testing our sources and others on a test stand, one which we are building at UChicago. We assembled the test stand with an existing electron source and performed initial thermal and vacuum tests.

In the future, to confirm the electron source designs, the geometries of the electrodes in the sources will be altered. Particularly, the electrode sizes and positions will be optimized for an appropriate beam distribution and current for both sources separately. Currently we have a design that would operate well for the current density requirements of the strong electron source. Next, we will further make changes to the design to meet the goals of the basic source. Once these sources are designed and their operations confirmed in TRAK, they will be manufactured, tested on the test stand, and finally commissioned in the IOTA ring.

The next commissioning steps for the test stand include improving the ratio pyrometer setup, testing the YAG:Ce screen and Fraday plate, as well as adding further features to the data acquisition user interface.

## ACKNOWLDGEMENTS

We would like to thank the teams at the University of Chicago, Fermilab, and the IOTA/FAST collaboration for their support. This research is funded by the NSF Graduate Research Fellowships Program (GRFP). The project is a collaboration between the University of Chicago and Fermilab. This manuscript has been authored by Fermi Research Alliance, LLC under Contract No. DE-AC02-07CH11359 with the U.S. Department of Energy, Office of Science, Office of High Energy Physics.